\RequirePackage{iftex}
\ifpdf
\PassOptionsToPackage{pdftex}{graphicx}
\else
\PassOptionsToPackage{dvipdfmx}{hyperref}
\PassOptionsToPackage{dvipdfmx}{xcolor}
\fi
\documentclass[fp,twocolumn]{jpsj3}
\usepackage{txfonts}%
\makeatletter
\providecommand\ext@figure{lof}
\providecommand\ext@table{lot}
\newsavebox{\jpsj@tabbox}
\newenvironment{ruledtabular}{\par\centering\small\let\jpsj@tabular\tabular\let\jpsj@endtabular\endtabular\def\tabular##1{\jpsj@tabular{##1}\hline\hline}\def\endtabular{\hline\hline\jpsj@endtabular}\begin{lrbox}{\jpsj@tabbox}}{\end{lrbox}\ifdim\wd\jpsj@tabbox>\linewidth\typeout{ruledtabular scaled to the column width: natural width \the\wd\jpsj@tabbox, line \the\inputlineno}\resizebox{\linewidth}{!}{\usebox{\jpsj@tabbox}}\else\usebox{\jpsj@tabbox}\fi\par}

\makeatother

\usepackage{dcolumn}%
\usepackage{bm}%
\usepackage{etoolbox}
\usepackage{xr-hyper}%
\usepackage{hyperref}%
\usepackage{cleveref}
\hypersetup{hidelinks}
\makeatletter
\let\jpsj@bibcite\bibcite
\def\bibcite#1#2{\jpsj@bibcite{#1}{#2}\global\@namedef{jpsj@key@#2}{#1}}
\renewcommand\citeform[1]{\@ifundefined{jpsj@key@#1}{#1}{\hyper@@link[cite]{}{cite.\@nameuse{jpsj@key@#1}}{\textcolor{jpsjlink}{#1}}}}
\def\@bibitem#1{\@skiphyperreftrue\H@item\@skiphyperreffalse\Hy@raisedlink{\hyper@anchorstart{cite.#1}\relax\hyper@anchorend}\if@filesw\begingroup\let\protect\noexpand\immediate\write\@auxout{\string\bibcite{#1}{\the\value{\@listctr}}}\endgroup\fi\ignorespaces}
\makeatother

\makeatletter
\def\@citess#1{\textsuperscript{\textcolor{jpsjlink}{#1\citeright}}}
\makeatother
\usepackage{xspace}
\crefname{section}{Sect.}{Sects.}
\Crefname{section}{Section}{Sections}
\crefname{subsection}{Sect.}{Sects.}
\Crefname{subsection}{Section}{Sections}
\crefname{subsubsection}{Sect.}{Sects.}
\Crefname{subsubsection}{Section}{Sections}
\crefname{equation}{Eq.}{Eqs.}
\Crefname{equation}{Equation}{Equations}
\crefname{figure}{Fig.}{Figs.}
\Crefname{figure}{Figure}{Figures}
\crefname{table}{Table}{Tables}
\Crefname{table}{Table}{Tables}
\crefname{appendix}{Appendix}{Appendices}
\Crefname{appendix}{Appendix}{Appendices}
\crefformat{appendix}{#2Appendix#3}
\Crefformat{appendix}{#2Appendix#3}
\makeatletter
\if@twocolumn
\AtBeginEnvironment{acknowledgments}{\footnotesize\paragraph{\footnotesize Acknowledgments}}
\fi
\makeatother
\newcommand{\jpsjemails}{\noindent$^{*}$koji{\_}kawahara{\_}aa@mail.toyota.co.jp\par
\noindent\phantom{$^{*}$}ORCID: \href{https://orcid.org/0000-0002-0596-0743}{0000-0002-0596-0743}\par
\noindent$^{\dagger}$kawashima@issp.u-tokyo.ac.jp\par
\noindent\phantom{$^{\dagger}$}ORCID: \href{https://orcid.org/0000-0002-8440-6037}{0000-0002-8440-6037}\par
\noindent$^{\ddagger}$hiroyuki{\_}suto@mail.toyota.co.jp\par
\noindent\phantom{$^{\ddagger}$}ORCID: \href{https://orcid.org/0009-0009-5266-3002}{0009-0009-5266-3002}\par}
\makeatletter
\patchcmd{\thebibliography}{\list}{\jpsjemails\list}{}{\typeout{jpsj e-mail patch of thebibliography failed}}
\makeatother
\makeatletter
\patchcmd{\balancing@outputdblcol}{\vskip\@interlinegap}{\vskip-15pt\relax}{}{\typeout{jpsj column-join patch failed}}
\makeatother

\usepackage{diffcoeff}
\usepackage{physics2}
\usephysicsmodule{ab, ab.braket}
\usepackage{xcolor}
\definecolor{jpsjlink}{HTML}{0000EE}
\usepackage[normalem]{ulem}

\newif\ifshowclaims
\showclaimsfalse
\makeatletter
\ifshowclaims
\colorlet{mk-gen-4}{blue!42}\colorlet{mk-gen-3}{blue!26}\colorlet{mk-gen-2}{blue!13}\colorlet{mk-gen-1}{blue!5}
\colorlet{mk-mod-4}{green!55}\colorlet{mk-mod-3}{green!35}\colorlet{mk-mod-2}{green!18}\colorlet{mk-mod-1}{green!7}
\colorlet{mk-num-4}{yellow!95!orange}\colorlet{mk-num-3}{yellow!60}\colorlet{mk-num-2}{yellow!32}\colorlet{mk-num-1}{yellow!12}
\colorlet{mk-con-4}{orange!55}\colorlet{mk-con-3}{orange!35}\colorlet{mk-con-2}{orange!18}\colorlet{mk-con-1}{orange!8}
\colorlet{mk-sup-4}{gray!60}\colorlet{mk-sup-3}{gray!40}\colorlet{mk-sup-2}{gray!22}\colorlet{mk-sup-1}{gray!10}
\colorlet{mk-app-4}{violet!50}\colorlet{mk-app-3}{violet!32}\colorlet{mk-app-2}{violet!16}\colorlet{mk-app-1}{violet!6}
\AtBeginDocument{\NewCommandCopy\mk@cref\cref\NewCommandCopy\mk@Cref\Cref\NewCommandCopy\mk@cite\cite\NewCommandCopy\mk@citen\citen}
\newcommand{\mk@citebox}[1]{\def\mk@citearg{#1}\futurelet\mk@next\mk@citeboxa}
\newcommand{\mk@citeboxa}{\let\mk@do\mk@citeboxn\ifx\mk@next.\let\mk@do\mk@citeboxp\fi\ifx\mk@next,\let\mk@do\mk@citeboxp\fi\ifx\mk@next;\let\mk@do\mk@citeboxp\fi\ifx\mk@next:\let\mk@do\mk@citeboxp\fi\mk@do}
\newcommand{\mk@citeboxn}{\mbox{\expandafter\mk@cite\expandafter{\mk@citearg}}}
\newcommand{\mk@citeboxp}[1]{\mbox{\expandafter\mk@cite\expandafter{\mk@citearg}#1}}
\newcommand{\mk@protect}{\protected\def\cref##1{\mbox{\mk@cref{##1}}}\protected\def\Cref##1{\mbox{\mk@Cref{##1}}}\protected\def\cite##1{\UL@stop\unpenalty\unskip\unpenalty\unskip\unkern\unskip\UL@start\mk@citebox{##1}}\protected\def\citen##1{\mbox{\mk@citen{##1}}}}
\newsavebox{\mk@wave}
\sbox{\mk@wave}{\usefont{U}{lasy}{m}{n}\fontsize{6}{6}\selectfont\char58}
\newcommand{\mk}[3]{\bgroup\mk@protect\markoverwith{\textcolor{mk-#1-#2}{\rule[-0.62ex]{1pt}{2.55ex}}}\ULon{#3}}
\newcommand{\mkr}[3]{\bgroup\mk@protect\markoverwith{\rlap{\textcolor{mk-#1-#2}{\rule[-0.62ex]{\wd\mk@wave}{2.55ex}}}\lower3.2pt\hbox{\textcolor{red}{\usebox\mk@wave}}}\ULon{#3}}
\newcommand{\rw}[1]{\bgroup\mk@protect\markoverwith{\lower3.2pt\hbox{\textcolor{red}{\usebox\mk@wave}}}\ULon{#1}}
\else
\newcommand{\mk}[3]{#3}
\newcommand{\mkr}[3]{#3}
\newcommand{\rw}[1]{#1}
\fi
\makeatother

\newif\ifshowchanges
\ifshowclaims\showchangestrue\else\showchangesfalse\fi
\makeatletter
\newif\ifchg@on
\definecolor{chgform}{RGB}{0,120,40}
\gdef\chg@col{blue}
\ifshowchanges
\patchcmd{\UL@start}{\bgroup\kern-3sp}{\bgroup\ifchg@on\color{\chg@col}\fi\kern-3sp}{}{\typeout{change-mark patch of UL@start failed}}
\DeclareRobustCommand{\chgon}{\global\chg@ontrue\gdef\chg@col{blue}\color{blue}}
\DeclareRobustCommand{\chgfon}{\global\chg@ontrue\gdef\chg@col{chgform}\color{chgform}}
\DeclareRobustCommand{\chgoff}{\global\chg@onfalse\color{black}\relax}
\else
\DeclareRobustCommand{\chgon}{}
\DeclareRobustCommand{\chgfon}{}
\DeclareRobustCommand{\chgoff}{}
\fi
\makeatother

\newcommand{\GdFeO}{GdFeO$_3$\xspace}
\newcommand{\NnDW}{NnDW\xspace}
\newcommand{\NPDW}{NPDW\xspace}
\newcommand{\nPDW}{nPDW\xspace}
\newcommand{\hk}{\hat\kappa}

\makeatletter
\def\ps@headings{\let\@mkboth\@gobbletwo\let\@oddhead\@empty\let\@evenhead\@empty
\def\@oddfoot{\hfill{\footnotesize\thepage}\hfill}\let\@evenfoot\@oddfoot}
\let\ps@jpsj\ps@headings
\makeatother
\begin{document}

\title{Curvature-Dependent Splitting of Field-Driven Multiferroic Domain Walls:\\
Where a Meandering Wall Splits First}

\author{Koji Kawahara$^{1*}$, Naoki Kawashima$^{2,3\dagger}$, and Hiroyuki Suto$^{1\ddagger}$}
\inst{$^{1}$Frontier Research Center, Toyota Motor Corporation, Susono, Shizuoka 410-1193, Japan\\
$^{2}$Institute for Solid State Physics, The University of Tokyo, Kashiwa, Chiba 277-8581, Japan\\
$^{3}$Trans-scale Quantum Science Institute, The University of Tokyo, Tokyo 113-0033, Japan}

\abst{%
\mk{con}{2}{\chgon In multiferroics, domain walls of coupled order parameters (constituent domain walls) bind into a composite domain wall,
which can split under a strong external field.\chgoff}
\mk{con}{3}{\chgon We investigate how curvature affects the splitting threshold and selects the splitting location.\chgoff}
\mk{gen}{4}{\chgon Analytically, for relaxational time-dependent Ginzburg--Landau equations to first order in curvature,
we show that curvature drives the constituent domain walls relative to each other
if and only if their stiffness-to-friction ratios differ among the order parameters.\chgoff}
\mk{num}{4}{\chgon In one-dimensional simulations of a \GdFeO-type model at fixed weak curvature,
curvature raises the \chgfon magnitude of the\chgoff{}\chgon{} threshold electric field when curvature and field push the wall the same way
and lowers it when they push oppositely.\chgoff}
\mk{num}{4}{\chgon In two-dimensional simulations, a meandering wall driven by a uniform electric field first splits
at the apex where the curvature drive opposes the field drive most strongly.\chgoff}
\mk{num}{3}{\chgon Overall, domain wall splitting is explained by the loss of a bound translating solution
and interpreted in terms of the balance between the relative driving force, which depends on curvature, and the binding force.\chgoff}
}

\maketitle

\section{\label{sec:intro}Introduction}

\begin{figure}[t]
\centering
\includegraphics{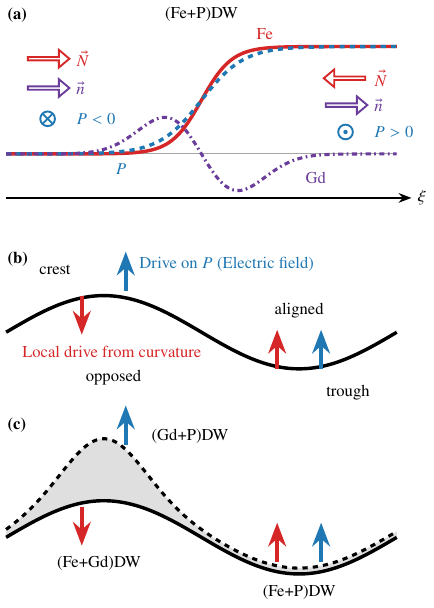}
\caption{Concept of curvature-induced splitting of a composite domain wall ($\vec N$, $\vec n$: Néel vectors of Fe and Gd; $P$: polarization).
(a) \chgfon Schematic profiles of a bound (Fe+P)DW (the \NPDW{} of \cref{sec:model-props});\chgoff{} hollow arrows: $\vec N$ and $\vec n$ in the two bulk states.
(b) Meandering wall: the field drives $P$, \chgfon whereas\chgoff{} curvature drives Fe.
(c) The crest splits into \chgfon an (Fe+Gd)DW and a (Gd+P)DW (\NnDW{} and \nPDW),\chgoff{} with a third bulk state (hatched) between \chgfon them. The splitting does not start at the trough.\chgoff}
\label{fig:concept}
\end{figure}

\mk{con}{2}{\chgon Systems with several coupled order parameters occur in many kinds of materials.
One example is multiferroic materials, in which the coupling between magnetization and electric polarization enables
electric-field control of magnetization~\cite{Chu2008,Heron2014,Fiebig2016,Fiebig2002}.
Other examples are liquid-crystal--colloid mixtures~\cite{Matsuyama2008}
\chgfon and magnetic multilayers\chgoff{} whose layers are coupled to each other~\cite{Fullerton1998,Metaxas2010}.
\chgfon In such systems, the coupling between the orders binds the domain walls of the individual order parameters (constituent domain walls) into a composite domain wall,
which has a complex internal structure [\cref{fig:concept}(a)].\chgoff{}\chgon{}
This raises a broad question of how the correlation among the coupled orders affects the motion of domain walls.\chgoff}

\mk{con}{1}{\chgon The static structure of domain walls of coupled order parameters has been described by Landau theory~\cite{Houchmandzadeh1991,Daraktchiev2010}.
In magnetic systems, the motion of coupled magnetic and ferroelectric domain walls under an electric field, and their decoupling,
have been studied in experiments and in micromagnetic calculations~\cite{Tokunaga2012,Skumryev2011,VandeWiele2016,Kagawa2009}.\chgoff}
\mk{con}{2}{\chgon Taking the magnetic multiferroic GdFeO$_3$ as an example, a previous study~\cite{OurPrevWork} showed that a flat composite domain wall
splits into two under a strong magnetic field, even in a pure system without impurities.
That study used a time-dependent Ginzburg--Landau (TDGL) model based on the experiment of Tokunaga \textit{et al.}~\cite{Tokunaga2009}.\chgoff}

\mk{con}{1}{Real domain walls, however, are generally curved:
shrinking domains are bounded by curved walls that move with a normal velocity
proportional to the local curvature~\cite{AllenCahn1979,Bray1994}, and walls bow out between pinning centers~\cite{Lemerle1998}.
\chgon Landau theory also gives static meandering domain walls of coupled antiferrodistortive and polar order parameters~\cite{Eliseev2019}.\chgoff}
\mk{con}{3}{\chgfon In this paper we investigate\chgoff{} how wall
curvature changes the threshold field \chgfon for the splitting of a composite domain wall and how it selects where the wall splits.\chgoff}

\mk{con}{1}{\chgfon Related phenomena have been studied along two lines of work.
The first concerns how walls and fronts bind and split.\chgoff{}
\chgon In one-dimensional TDGL models whose free energy has three local minima, a front between the disordered and the stable phase can split into two fronts enclosing a metastable phase~\cite{BechhoeferLowenTuckerman1991}.
With two nonconserved order parameters of equal relaxation times and equal gradient coefficients, the steady front can disappear at a saddle-node bifurcation~\cite{TuckermanBechhoefer1992}.\chgoff{}
\chgfon Binding or separation of walls, fronts, and solitons has also been studied in
propagating terraces of reaction--diffusion fronts~\cite{FifeMcLeod1977,Risler2008,DucrotGilettiMatano2014},
exchange springs~\cite{Goto1965,Fullerton1998,Kronmuller1987,KronmullerGoll2002},
coupled magnetic layers~\cite{Metaxas2010},
bound soliton pairs~\cite{Condat1983,Campbell1986,Kivshar1989},
two-component condensates~\cite{Charalampidis2015,Alotaibi2017},
and liquid-crystal walls~\cite{deLozar2005,Yi2024}.\chgoff{}}

\mk{con}{1}{\chgfon The second line concerns how curvature changes the motion of fronts with several components,
beyond the curvature-driven motion of interfaces of a single order parameter.\chgoff{}
\chgon For weak curvature, the normal velocity of a two-component reaction--diffusion front differs from the planar-front velocity
by an amount proportional to the curvature~\cite{Meron1992}.
In a two-component model near a nonequilibrium Ising--Bloch bifurcation, curvature also acts through a coefficient
that vanishes when the two components have equal diffusion coefficients.\chgoff{}
\chgfon There, negative curvature can locally nucleate
a front segment that propagates in the reverse direction~\cite{HagbergMeron1997,HagbergMeron1998}.\chgoff}

\mk{con}{2}{\chgon The studies above have treated how walls and fronts bind and split, including the splitting of a flat composite domain wall in GdFeO$_3$,
and how curvature changes front motion.
To our knowledge, however, how curvature changes the threshold of such splitting, and where along a curved wall the splitting starts, have not been examined.
We examine both for an electric-field-driven composite domain wall in a \GdFeO-type model.
We also interpret the splitting semi-analytically as a loss of balance between a curvature-dependent relative driving force on the constituent domain walls and the binding force between them.\chgoff{}}

\mk{con}{2}{\chgfon The rest of this paper is organized as follows.\chgoff}
\mk{gen}{4}{\chgon \Cref{sec:general} treats relaxational TDGL equations~\cite{HohenbergHalperin1977} to first order in weak curvature.
We show that curvature drives the constituent domain walls relative to each other
if and only if their stiffness-to-friction ratios $\kappa_X/\Gamma_X$ (gradient coefficient divided by friction coefficient) differ among the order parameters $X$.
For a one-dimensional wall, this result gives a necessary condition for curvature to change the threshold field.\chgoff}

\mk{mod}{2}{\chgon \Cref{sec:model} introduces a \GdFeO-type model and the numerical setup.
In this model, $\kappa_X/\Gamma_X$ differs among the antiferromagnetic Néel vectors of Fe and Gd and the polarization $P$,
the electric field acts only on $P$, and curvature acts mainly on Fe [\cref{fig:concept}(b)].\chgoff}

\mk{num}{4}{\chgon \Cref{sec:phase} gives one-dimensional results for this model at fixed weak curvature.
The magnitude of the threshold electric field for splitting (hereafter, the threshold field) increases relative to its flat-wall value
where curvature and field push the wall in the same direction (aligned).
It decreases where they push it in opposite directions (opposed).
At stronger curvature, curvature alone can split the wall.
\chgon We interpret the threshold field at two levels of approximation.\chgoff{}
\chgfon At the first level, the threshold field coincides with the largest field at which a boundary-value problem has a bound translating solution.\chgoff{}\chgon{}
At the second level, a collective-coordinate model, which reduces the wall to the distance between its constituent domain walls, balances a curvature-dependent relative driving force on these constituent domain walls against the binding force between them.
For the main parameters at weak drive, it reproduces the flat-wall threshold field \chgfon of the simulations,\chgoff{}\chgon{} the slope of its change with curvature on the aligned side, and the curvature at which the threshold field on the opposed side vanishes, each \chgfon to within several percent of the simulated value.\chgoff{}
\chgon At larger curvature, we compare the boundary with the largest curvature at which one of the two composite domain walls produced by the splitting remains bound.\chgoff}

\mk{num}{4}{\chgon \Cref{sec:2d} gives two-dimensional results.
A meandering wall driven by a uniform field splits first
at the apex where curvature and field push the wall in opposite directions (the crest) [\cref{fig:concept}(c)], not at the opposite apex (the trough).
The split region then extends along the wall toward the trough.
In the weak-curvature range, we interpret this selection in terms of the apex curvature and its time evolution.\chgoff}

\mk{con}{2}{\chgon \Cref{sec:discussion} discusses possible experimental tests on \GdFeO and on other similar systems, together with the limitations of this study. \Cref{sec:conclusion} concludes.
The \cref{app:reduced-model} \chgfon gives the coefficients and thresholds of the collective-coordinate model,\chgoff{}\chgon{} and the Supplemental Material~\cite{SuppMat} (Supp.) provides
further details.\chgoff}

\section{\label{sec:general}General Theory of Multi-Order TDGL with Finite Curvature}

\subsection{\label{sec:assumptions}Equations of motion and definition of splitting}

\mk{gen}{2}{We represent the order parameters by real fields $\phi_X$ ($X=1,\dots,n$) and take the free energy}
\begin{equation}
\mathcal{F}=\int d^2r\,\ab[\sum_X\frac{\kappa_X}{2}(\nabla\phi_X)^2+f(\phi_1,\dots,\phi_n)],
\label{eq:F-general}
\end{equation}
\mk{gen}{2}{where $\kappa_X\ge0$ is the stiffness and $f$ \chgfon is\chgoff{} the local free-energy density.
A spatially uniform external field may couple, through $f$, to some of the order parameters.
Each order parameter obeys, with friction $\Gamma_X>0$,}
\begin{equation}
\Gamma_X\,\partial_t\phi_X=-\frac{\delta\mathcal{F}}{\delta\phi_X}=\kappa_X\nabla^2\phi_X-\frac{\partial f}{\partial\phi_X}.
\label{eq:tdgl-general}
\end{equation}
\mk{gen}{2}{\chgon The order parameters are assumed to be nonconserved, and noise is neglected.
We also assume that they couple to one another and to the field only through $f$, which contains no spatial derivatives.\chgoff}

\mk{gen}{3}{We define the splitting of a single composite domain wall as follows.
Choose two order parameters $X$ and $Y$ that change value across the
composite domain wall \chgfon[e.g., Fe and $P$ in \cref{fig:concept}(a)]\chgoff, and let $d(t)$ be the distance between the center of constituent domain wall $X$ and \chgfon that of constituent domain wall $Y$.
The center of a constituent domain wall is the position where its order parameter takes the value midway between its two bulk values.\chgoff{}
\chgfon We say that the wall splits when $d(t)$ grows without bound.\chgoff} \mk{gen}{2}{In contrast, $d$ settles to a constant value for a wall that approaches a bound translating solution,
in which all constituent domain walls translate \chgfon as a single composite domain wall\chgoff{} with a common constant velocity and keep their shape.}
\mk{gen}{3}{For a two-dimensional wall, we use the distance $d(s,t)$ at each position $s$ along the wall
\chgfon and say that the wall splits at $s$ when $d(s,t)$ grows without bound\chgoff{}
\chgfon(see \cref{sec:calc} for the criterion used in finite-time numerical calculations)\chgoff.}
\mk{gen}{2}{\chgfon Suppose that a third bulk state exists that differs from both end bulk states and is locally stable.
A split wall then becomes two spatially separated composite domain walls, with an expanding region of the third bulk state between them.
The two walls inherit different combinations of the original constituent domain walls.
The region between them must be in a bulk state, because otherwise its excess free energy would grow without bound as $d(t)$ increases
[see \cref{fig:split-def}(iii) for the model of \cref{sec:model}]\chgoff.}

\subsection{\label{sec:cross-section}Cross-section equation\texorpdfstring{ \chgfon of a curved domain wall\chgoff}{ of a curved domain wall}}

\mk{gen}{2}{For a domain wall in a two-dimensional system, we take curvilinear coordinates $(s,\xi)$ along the wall contour:
$s$ is the arc length along the contour and $\xi$ \chgfon is\chgoff{} the signed normal
coordinate. The curvature $K$ is positive when the center of curvature lies on the $-\xi$ side ($K=+\nabla\cdot\hat\xi$,
where $\hat\xi$ is the
unit normal in the $+\xi$ direction\chgfon; \cref{fig:sign-convention}\chgoff). \chgfon In these coordinates, the Laplacian is\chgoff}
\begin{equation}
\nabla^2=\frac{1}{1+\xi K}\,\partial_\xi\ab[(1+\xi K)\,\partial_\xi]
+\frac{1}{1+\xi K}\,\partial_s\ab[\frac{1}{1+\xi K}\,\partial_s].
\label{eq:laplacian-curv}
\end{equation}
\mk{gen}{2}{We assume that the \chgfon curvature $K(s)$ varies slowly along the wall, on the scale of\chgoff{} the total
\chgfon width\chgoff{} of
the composite domain wall (including the widths of the constituent domain walls and the distances between \chgfon them).
Then the cross section changes adiabatically along the wall.
We also assume that the curvature itself is small, with a radius of curvature $1/|K(s)|$ much larger than this width,
and\chgoff{} treat the cross-section equation to first order in \chgfon curvature.\chgoff}
\mk{gen}{2}{Dropping the derivatives \chgfon with respect to\chgoff{} $s$ and expanding to first order in $\xi K$ gives}
\begin{equation}
\partial_\xi^2 \ \longrightarrow\ \partial_\xi^2+K(s)\,\partial_\xi+O\ab(K^2 \xi)\,\partial_\xi.
\label{eq:curv-drift}
\end{equation}
\mk{gen}{2}{The cross-section equation of each order parameter is therefore}
\begin{equation}
\Gamma_X\,\partial_t\phi_X=\kappa_X\ab(\partial_\xi^2\phi_X+K\,\partial_\xi\phi_X)-\frac{\partial f}{\partial\phi_X}.
\label{eq:cross-section}
\end{equation}

\mk{gen}{2}{Assume that the composite domain wall stays \chgfon bound and moves with velocity $v$ (positive in the $+\xi$ direction).\chgoff{}
Switching to the comoving frame
$\zeta=\xi-vt$, with $\partial_t\to\partial_t-v\,\partial_\zeta$, \cref{eq:cross-section} becomes}
\begin{equation}
\Gamma_X\,\partial_t\phi_X=\kappa_X\partial_\zeta^2\phi_X+\ab(\kappa_XK+\Gamma_Xv)\,\partial_\zeta\phi_X-\frac{\partial f}{\partial\phi_X}.
\label{eq:comoving-shift}
\end{equation}

\subsection{\label{sec:curvature-drive}Order-parameter-dependent curvature drive}

\mk{gen}{4}{Proposition~1 (for \cref{eq:cross-section} with $K\neq0$).
A change to a frame $\zeta=\xi-v_0t$ with a common constant velocity $v_0$ removes the curvature terms
from the cross-section equations of all order parameters simultaneously if and only if}
\begin{equation}
\kappa_X=\lambda\,\Gamma_X,\quad\forall X,
\label{eq:prop1}
\end{equation}
\mk{gen}{4}{\chgfon in which case\chgoff{} $v_0=-\lambda K$.}
\mk{gen}{3}{This is exact for the cross-section equation truncated at first order in curvature.
\chgon A common ratio $\kappa_X/\Gamma_X$ corresponds to equal diffusion coefficients.
In that case, Proposition~1 reduces to a known result for reaction--diffusion fronts.
The normal velocity of a curved front then differs from the planar-front velocity
by the diffusion coefficient times the local curvature, even without a separation of time scales~\cite{Meron1992}.
Proposition~1 states the condition in general form, for any number of relaxational fields and any local free energy, to first order in curvature.\chgoff}

\mk{gen}{4}{Corollary~1 (one dimension, \chgfon weak curvature\chgoff; the system is infinite \chgfon in the direction normal to the wall, with stable bulk states on both sides).\chgoff{}
If \cref{eq:prop1} holds, \chgfon the curvature affects neither whether the wall splits nor the threshold field,
both for a fixed $K$ and for a time-dependent\chgoff{} $K(t)$.}
\mk{gen}{3}{Indeed, if $\phi_X^0(\xi,t)$ solves the flat-wall problem,
$\phi_X^0(\xi+\lambda\int_0^tK\,dt',t)$ solves the curved one
with the same boundary conditions, and this uniform translation leaves $d(t)$ unchanged.}
\mk{gen}{3}{Within the scope of Corollary~1, curvature can therefore change whether the wall splits, or its threshold field,
only if \cref{eq:prop1} is violated.
\chgon The front-splitting models~\cite{BechhoeferLowenTuckerman1991,TuckermanBechhoefer1992} have either a single order parameter or two order parameters with equal relaxation times and equal gradient coefficients, and thus satisfy \cref{eq:prop1}.
By Corollary~1, curvature does not change their splitting in one dimension at first order in curvature.\chgoff{}
\chgfon A violation of \cref{eq:prop1}, although necessary, is not sufficient for splitting by curvature alone at zero field.\chgoff}

\mk{gen}{2}{\chgfon Assume that $K$ is constant in time (an assumption used only in the one-dimensional analysis) and that a bound translating solution
exists in the infinite system of Corollary~1.
Multiplying the steady form of \cref{eq:comoving-shift} by $\partial_\zeta\phi_X$, summing over $X$, and integrating over $\zeta$
then yields\chgoff{} the comoving power-balance identity}
\begin{equation}
v\sum_X\Gamma_XD_X=-K\sum_X\kappa_XD_X+\ab[f]_{-\infty}^{+\infty},
\label{eq:power-balance-main}
\end{equation}
\mk{gen}{2}{where $D_X\equiv\int_{-\infty}^{\infty}(\partial_\zeta\phi_X)^2d\zeta$ is the dissipation weight of $\phi_X$
and $[f]_{-\infty}^{+\infty}$ is the difference between the free-energy densities of the two end bulk \chgfon states.
The\chgoff{} identity is exact for \cref{eq:comoving-shift} with any $f$, $\kappa_X$, and \chgfon$\Gamma_X$.\chgoff}
\mk{gen}{2}{The velocity of the whole wall therefore obeys the implicit relation
($D_X$ depends on the unknown \chgfon spatial profile of $\phi_X$\chgoff{} and on $v$)}
\begin{equation}
v=-\lambda^*K+\frac{\ab[f]_{-\infty}^{+\infty}}{\sum_X\Gamma_XD_X},\quad
\lambda^*\equiv\frac{\sum_X\kappa_XD_X}{\sum_X\Gamma_XD_X},
\label{eq:velocity-whole}
\end{equation}
\mk{gen}{2}{where $\lambda^*$ is the effective tension divided by the effective friction ($\lambda^*=\lambda$ under \cref{eq:prop1}).}
\mk{gen}{3}{In the frame moving at this $v$, the first-derivative coefficient of \cref{eq:comoving-shift} becomes
$K(\kappa_X-\lambda^*\Gamma_X)+\Gamma_X\ab[f]_{-\infty}^{+\infty}/\sum_Y\Gamma_YD_Y$.
Its first term is the relative drive by curvature on constituent domain wall $X$,
measured in the frame of the whole composite domain wall.
\chgon It vanishes for every $X$ when \cref{eq:prop1} holds and otherwise differs among the order parameters.
The second term is the friction $\Gamma_X$ times a velocity common to all $X$, and by itself it only translates the wall as a whole.\chgoff}
\mk{gen}{4}{Heuristically, splitting is expected when the \chgfon coupling between the order parameters\chgoff{} can no longer balance \chgfon the combined effect of this relative drive
and of the part of the field drive that acts on only some of the order parameters.\chgoff{}
\chgfon The\chgoff{} wall then no longer approaches a bound translating solution, and \cref{eq:power-balance-main} no longer \chgfon applies.
The force balance at weak drive is made quantitative in \cref{sec:theory-check}.\chgoff}
\mk{sup}{1}{\chgfon The assumptions of the identity and its form for the model of \cref{sec:model} are stated in Supp.~\cref{supp-supp:general-identity}.
Supp.~\cref{supp-supp:z2cubed} treats a model with three coupled Z$_2$ fields at zero field.
It gives an example of a wall that curvature alone splits and an example of a wall that stays bound\chgoff{}\chgon, both with \cref{eq:prop1} violated.\chgoff}

\section{\label{sec:model}Model and Numerical Method}

\mk{mod}{2}{\chgon As a concrete example of a multiferroic that violates \cref{eq:prop1}, we consider a TDGL model of \GdFeO type in the following.\chgoff}

\subsection{\label{sec:tdgl}\GdFeO-type TDGL model}

\mk{mod}{2}{The model is based on the Ginzburg--Landau free energy of our previous work~\cite{OurPrevWork}
for the antiferromagnetic order parameters of Fe and Gd\chgfon,\chgoff{} and the polarization.
\chgfon In contrast to that work,\chgoff{} the magnitudes of the Néel vectors are fixed,
the polarization has no stiffness term, and the external field \chgfon is an electric field that couples to $P$ rather than a field that couples to\chgoff{} the Fe Néel vector.}
\mk{mod}{2}{\chgfon We consider temperatures well below the Néel temperatures of Gd and Fe and fix the magnitudes of the Néel vectors $\vec N$ (Fe)
and $\vec n$ (Gd) at\chgoff{} $|\vec N|=S_{\rm Fe}=5/2$ and $|\vec n|=S_{\rm Gd}=7/2$.
The free-energy density is}
\begin{equation}
\begin{split}
f_{\rm tot}={}&\frac{\kappa_F}{2}(\nabla\vec N)^2-\frac{\alpha_x}{2}(\vec N\cdot\vec e_x)^2
+\frac{\alpha_z}{2}(\vec N\cdot\vec e_z)^2\\
&+\frac{\kappa_G}{2}(\nabla\vec n)^2+\frac{A}{2}P^2-EP+cP\,\vec N\cdot\vec n,
\end{split}
\label{eq:F-model}
\end{equation}
\mk{mod}{2}{and the free-energy functional is \chgfon$\mathcal{F}_{tot}=\int d^2r\,f_{\rm tot}$.\chgoff{}
The spin axes are $x$ (easy), $y$ (\chgfon intermediate\chgoff), and $z$ (hard), the last being the crystal $c$ axis of \GdFeO;
$P$ and the electric field $E$ point along $z$.
The subscripts of $\vec e_x$ and $\vec e_z$ label spin axes, not the real-space coordinates used for the wall.
\chgfon The coefficients\chgoff{} $\kappa_{F,G}$ are exchange \chgfon stiffnesses, and\chgoff{} $\alpha_x,\alpha_z>0$ are on-site anisotropies of \chgfon Fe.
The Néel vector $\vec N$ thus prefers to rotate in the
$xy$ plane at a domain wall.
The coefficient\chgoff{} $A>0$ is the inverse dielectric \chgfon susceptibility, and\chgoff{} $c>0$ is the magnetoelectric coupling coefficient, thought to
originate from a uniform displacement of Gd ions along the crystal $c$ axis due to exchange striction~\cite{Tokunaga2009}.}
\mk{mod}{3}{Gd is taken to be isotropic, and the stiffness term of the polarization is ignored ($\kappa_P=0$) because it is weak
compared with the stiffness terms of the Fe and Gd \chgfon orders.\chgoff}

\mk{mod}{2}{A tilde marks coefficients with the fixed magnitudes absorbed ($\tilde\kappa_F=|\vec N|^2\kappa_F$,
$\tilde\kappa_G=|\vec n|^2\kappa_G$, $\tilde\alpha_x=|\vec N|^2\alpha_x$, $\tilde c=|\vec N||\vec n|c$), and
$p\equiv AP/\tilde c$ is the rescaled polarization.
Each order parameter obeys a purely relaxational equation of \chgfon motion, the TDGL equation (Model~A)~\cite{HohenbergHalperin1977},\chgoff}
\begin{equation}
\begin{gathered}
\Gamma_N\,\partial_t\hat N=-\mathcal{P}_{\hat N}\frac{\delta\mathcal{F}_{tot}}{\delta\hat N},\quad
\Gamma_n\,\partial_t\hat n=-\mathcal{P}_{\hat n}\frac{\delta\mathcal{F}_{tot}}{\delta\hat n},\\
\Gamma_P\,\partial_t p=-\frac{\delta\mathcal{F}_{tot}}{\delta p},
\end{gathered}
\label{eq:tdgl-vector}
\end{equation}
\mk{mod}{2}{where $\Gamma_N$, $\Gamma_n$, and $\Gamma_P$ are the \chgfon friction coefficients of Fe, Gd, and the polarization (the last defined
for $p$),\chgoff{} $\hat N=\vec N/|\vec N|$, $\hat n=\vec n/|\vec n|$, and $\mathcal{P}_{\hat u}=1-\hat u\hat u^{\mathsf T}$
projects onto the \chgfon plane perpendicular to the unit vector $\hat u$.\chgoff}

\mk{mod}{2}{In the units below, the \chgon model has seven independent parameters. They are\chgoff{} the stiffness ratio $\varepsilon$, the
coupling strength $g$, the friction ratios $\gamma_n$ and $\gamma_P$, the rescaled field $e$, \chgon the rescaled curvature $\hk$, and the anisotropy ratio $\alpha_z/\alpha_x$.
A solution that starts in the easy plane stays in it, and $\alpha_z$ determines only the stability of this plane against out-of-plane deviations (Supp.~\cref{supp-supp:polar,supp-supp:reduced-eqs}).
We fix $\alpha_z/\alpha_x=10$, and the wall dynamics in the easy plane is governed by the other six parameters.\chgoff}
\mk{mod}{2}{Lengths are measured in units of the Fe wall width without coupling,
$\Delta_F=\sqrt{\tilde\kappa_F/\tilde\alpha_x}$~\cite{SchryerWalker1974};
the rescaled coordinate normal to the wall is $z=\xi/\Delta_F$,
and the curvature is $\hk\equiv K\Delta_F$.
Time is measured in units of \chgfon$\tau_N\equiv\Gamma_N/\tilde\alpha_x$, the relaxation time of a small in-plane deviation of a uniform Fe Néel vector
from its easy axis. The wall speed is measured in units of $\Delta_F/\tau_N$, and energy densities in units of $\tilde\alpha_x$.\chgoff{}
\chgfon The rescaled field is $e\equiv E/\tilde c$. The stiffness ratio and the coupling strength are\chgoff}
\begin{equation}
\varepsilon = \frac{\tilde\kappa_G}{\tilde\kappa_F}, \qquad
g = \frac{\tilde{c}^2}{2A\tilde\alpha_x},
\label{eq:eps-g-def}
\end{equation}
\mk{mod}{2}{and the friction ratios \chgfon are\chgoff{} $\gamma_n\equiv\Gamma_n/\Gamma_N$ and $\gamma_P\equiv\Gamma_P/\Gamma_N$ (the Gd friction and
the polarization friction, each divided by the Fe friction).}
\mk{mod}{2}{\chgfon The friction $\Gamma_P$ of $p$ is related to the friction $\Gamma_P^{\rm ph}$ of the unrescaled $P$ by $\Gamma_P=(\tilde c/A)^2\,\Gamma_P^{\rm ph}$.\chgoff{}
\chgfon The ratio of the relaxation time $\tau_P$ of $P$ to $\tau_N$ is thus $\tau_P/\tau_N=\gamma_P/(2g)=\tilde\alpha_x\Gamma_P^{\rm ph}/(A\Gamma_N)$, which does not contain $\tilde c$. Varying $g$ through $\tilde c$ therefore keeps
$\tau_P/\tau_N$, and hence $\gamma_P/g$, fixed.\chgoff}

\mk{mod}{3}{\chgfon In the easy plane, we measure the azimuthal angles $\Phi$ and $\phi$ of $\vec N$ and $\vec n$ from the $x$ axis and
define the relative angle $\psi\equiv\Phi-\phi$. The equations of motion, with the curvature drift term of \cref{eq:curv-drift} included, then reduce to\chgoff}
\begin{equation}
\begin{aligned}
\partial_t\Phi&=\partial_z^2\Phi+\hk\,\partial_z\Phi-\tfrac12\sin2\Phi+2g\,p\sin\psi,\\
\gamma_n\,\partial_t\phi&=\varepsilon\ab(\partial_z^2\phi+\hk\,\partial_z\phi)-2g\,p\sin\psi,\\
\gamma_P\,\partial_t p&=-2g\ab(p+\cos\psi-e).
\end{aligned}
\label{eq:tdgl-dimless}
\end{equation}
\mk{mod}{3}{\chgfon These equations\chgoff{} have the form of \cref{eq:cross-section} for $X=\Phi,\phi,p$ with $(\kappa_X,\Gamma_X)=(1,1)$, $(\varepsilon,\gamma_n)$, and
$(0,\gamma_P)$\chgfon. The stiffness-to-friction ratio $\kappa_X/\Gamma_X$ thus differs among the order parameters, so \cref{eq:prop1} does not hold.\chgoff{}
\chgon The curvature drift is thus largest for Fe, smaller by the factor $\varepsilon/\gamma_n$ for Gd, and absent for $P$, whereas the field $e$ enters only the equation for $p$.
The simulations of \cref{sec:phase,sec:2d} integrate \cref{eq:tdgl-vector}. The reduced \cref{eq:tdgl-dimless}, which are exact for a solution that stays in the easy plane, are used in the collective-coordinate models and in the boundary-value problem of \cref{sec:theory-check}.\chgoff}
\mk{sup}{1}{\chgfon The reduction to the easy plane is derived in Supp.~\cref{supp-supp:polar,supp-supp:reduced-eqs}.
The parameter estimates, the SI units, and the relaxation time of $P$ are discussed in Supp.~\cref{supp-supp:params}.\chgoff}

\subsection{\label{sec:model-props}Bulk states, composite domain walls, and symmetry}

\begin{figure*}[t]
\begin{minipage}[t]{0.665\textwidth}
\centering
\includegraphics{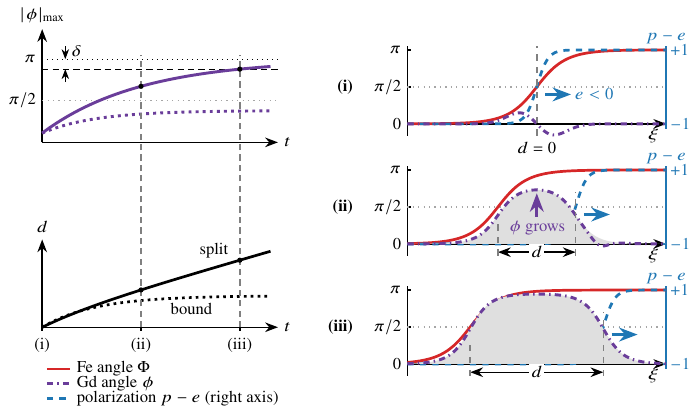}
\caption{Schematic of the splitting of \chgfon an\chgoff{} \NPDW{}\chgon{} for $e<0$\chgoff{}.
Left: $|\phi|_{\max}(t)$ and $d(t)$ for \chgfon a wall that splits and for a wall that stays bound;\chgoff{} \chgon horizontal dashed line: $\pi-\delta$;\chgoff{} (i)--(iii) mark the times of the
profiles on the right.
Right: profiles \chgfon as functions of $\xi$ (i) before the drive is switched on, (ii) during the splitting, and (iii) after the splitting,\chgoff{} with a third bulk
state between \chgfon the \NnDW (smaller $\xi$) and the \nPDW (larger $\xi$).\chgoff}
\label{fig:split-def}
\end{minipage}\hfill
\begin{minipage}[t]{0.315\textwidth}
\centering
\includegraphics{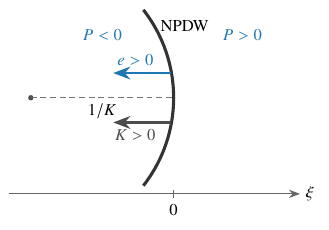}
\caption{\protect\raggedright Sign convention for \chgfon the\chgoff{} \NPDW.}
\label{fig:sign-convention}
\end{minipage}
\end{figure*}

\mk{mod}{2}{\chgon The Fe anisotropy allows $\Phi=0$ or $\pi$, and the coupling $cP\,\vec N\cdot\vec n$ holds Gd either parallel or antiparallel to Fe.
There are thus four bulk states,\chgoff{}
with $\Phi$ and $\phi$ each equal to $0$ or $\pi$ and $p=e-\cos\psi$\chgfon.
At\chgoff{} $E=0$ they are degenerate, and the sign of $P$ is opposite to that of $\vec N\cdot\vec n$.}
\mk{mod}{2}{\chgfon Three kinds of composite domain walls connect the bulk states. Across each of them, two of the three order
parameters change sign, so that $cP\,\vec N\cdot\vec n<0$ holds on both sides of the wall.\chgoff{}
We name \chgfon each wall after the pair of order parameters that reverses,\chgoff{} with N for the Fe Néel vector, n for the Gd Néel vector, P for the polarization, and DW
for domain wall. Across \chgfon an\chgoff{} \NnDW, $\Phi$ and $\phi$ both advance by $\pi$; across \chgfon an\chgoff{} \NPDW,
only $\Phi$ advances by $\pi$ [the (Fe+P)DW of \cref{fig:concept}(a)]; across \chgfon an\chgoff{} \nPDW, only $\phi$ advances
by $\pi$.} \mk{mod}{3}{\chgfon The bulk states on the two sides of an \NPDW have the same $\phi$. The third bulk state that can appear inside the wall when it splits has
$\phi$ shifted by $\pi$ from this value. An \NPDW therefore splits into an \NnDW and an \nPDW\chgoff{} with this third bulk state between them
[\cref{fig:concept}(c) and \cref{fig:split-def}(iii)].}
\mk{mod}{3}{\chgfon For\chgoff{} $|e|>1$ one of the two end bulk states of \chgfon the\chgoff{} \NPDW and of \chgfon the\chgoff{} \nPDW is not locally stable,
whereas for $|e|<1$ all four bulk states \chgfon are (Supp.~\cref{supp-supp:bulk-range}). Hereafter, we consider only $|e|<1$.\chgoff}

\mk{mod}{2}{Gd has no anisotropy of its own, so when $\psi$ changes, the only restoring term is the effective anisotropy
$\tilde c^2/A$ induced by the coupling through the polarization. Its balance with the Gd stiffness sets the width
$\Delta_G\equiv\sqrt{\tilde\kappa_GA}/\tilde c$ over which $\psi$ changes.
The ratio of $\Delta_G$ to the Fe wall width is $\Delta_G/\Delta_F=\sqrt{\varepsilon/(2g)}$\chgfon.
We refer to the regime $\Delta_G<\Delta_F$ as\chgoff{} the narrow-$\Delta_G$ side.}

\mk{mod}{2}{In this model, the term $\ab[f]_{-\infty}^{+\infty}$ of \cref{eq:power-balance-main} is
$2g\,e\ab[\hat N\cdot\hat n]_{-\infty}^{+\infty}$; its magnitude is $4g|e|$ for \chgfon an\chgoff{} \NPDW and \chgfon an\chgoff{} \nPDW and $0$
for \chgfon an\chgoff{} \NnDW.} \mk{mod}{3}{At $K=0$, the electric field therefore propels a bound translating \NPDW or
\nPDW but not \chgfon an\chgoff{} \NnDW.}

\mk{mod}{2}{For any parameters, the equations of motion with the drift term are exactly symmetric under the coordinate inversion
(in two dimensions, $x\to-x$ for the coordinate $x$ normal to the wall)}
\begin{equation}
(\xi,K)\ \longrightarrow\ (-\xi,-K),
\label{eq:sym-coord}
\end{equation}
\mk{mod}{2}{and, separately, under the field reversal (the subscripts are spin axes)}
\begin{equation}
(N_x,n_y,n_z,P,E)\ \longrightarrow\ (-N_x,-n_y,-n_z,-P,-E),
\label{eq:sym-field}
\end{equation}
\mk{mod}{2}{which \chgfon in\chgoff{} the easy plane reads $(\Phi,\phi,p)\to(\pi-\Phi,-\phi,-p)$. Each map alone exchanges the two end bulk states;
their product restores them and maps \chgfon the\chgoff{} \NPDW onto itself with $(K,E)\to(-K,-E)$.} \mk{mod}{3}{\chgfon The static \NPDW is
invariant under the product, because $\Phi-\pi/2$, $\phi$, and $p$ are odd functions of $\xi$ about the wall center.
The product therefore maps the initial state, $d$, and $|\phi|_{\max}$ (\cref{sec:calc}) onto themselves.
The results for $e>0$ thus follow from those for $e<0$ with the sign of $K$ reversed.\chgoff}

\subsection{\label{sec:calc}Sign convention, numerical \texorpdfstring{\chgfon setup\chgoff}{setup}, and splitting criterion}

\mk{mod}{2}{\chgfon Most of the calculations are for the \NPDW.\chgoff}
\mk{mod}{3}{Unless stated otherwise, \chgfon the \NPDW connects\chgoff{} $(\Phi,\phi,p)=(0,0,-1)$ at $\xi\to-\infty$ to $(\pi,0,+1)$ at
$\xi\to+\infty$.
\chgfon Then $e>0$ favors $p=+1$, the bulk state on the $+\xi$ side, and the field moves the wall toward $-\xi$ so that this domain grows.
A curvature $K>0$ places the center of curvature on the $-\xi$ side (\cref{sec:cross-section}), and the wall moves toward this center.
Each of the two therefore drives the wall toward $-\xi$, in accordance with the signs of the two terms on the right-hand side of
\cref{eq:power-balance-main} (\cref{fig:sign-convention}).\chgoff{}
Curvature and field are \emph{aligned} when they push the wall in the same direction ($Ke>0$) and
\emph{opposed} otherwise ($Ke<0$).
In a meandering wall under a uniform field, \chgfon the field moves the wall in one direction.
We call the apex that protrudes in this direction the crest, and the opposite apex the trough. The crest is opposed, and the trough is aligned\chgoff{} [\cref{fig:concept}(b)].}

\mk{mod}{2}{In both one and two dimensions, we solve \cref{eq:tdgl-vector} numerically \chgfon for five scalar fields (the polar and azimuthal angles of each Néel vector and $p$).\chgoff
\chgfon Each run starts from a static wall in the easy plane, and the drive is switched on as a step function at $t=0$.\chgoff}
\mk{mod}{2}{In one dimension, $\hk$ is \chgfon fixed\chgoff{} and enters through the drift term of \cref{eq:curv-drift}\chgfon. The computational domain\chgoff{} is
re-centered on the \chgfon Fe constituent domain wall as the wall moves\chgoff.
In two dimensions, the curvature arises from the wall shape itself\chgfon. The computational domain is periodic in $y$ (along the wall),
with period $L_y=135.8\,\Delta_F$ in the runs shown in \cref{sec:2d}, and has Neumann boundary conditions in $x$ (normal to the wall).
These are imposed at least $28\,\Delta_F$ away from the initial apexes of the meander,\chgoff{}\chgon{} which leaves enough space for the order parameters to reach their bulk values well before the boundaries.\chgoff}

\mk{mod}{3}{For \chgfon the\chgoff{} \NPDW{} in this model,
the distance $d$ of \cref{sec:assumptions} is taken between the center of Fe \chgfon($\Phi=\pi/2$) and the center of the polarization, and splitting is\chgoff{}
judged from whether $d$ \chgfon saturates\chgoff{} or keeps growing (\cref{fig:split-def}, left).
\chgon Numerically, we detect the onset of splitting by the criterion that\chgoff{}\chgfon{} the maximum deviation $|\phi|_{\max}$ of Gd from its bulk direction
reaches $\pi-\delta$, with $\delta=1^\circ$. Inside the third bulk state this deviation approaches $\pi$.
The time at which $|\phi|_{\max}$ reaches $\pi-\delta$ is $t_{\rm split}$.
The non-splitting criterion is that $|\phi|_{\max}$ and $d$ approach constant values
or that the wall starts to relax back to a bound \NPDW.\chgoff{}
\chgon A run that meets neither criterion within its time limit is counted as non-splitting (Supp.~\cref{supp-supp:num-criteria,supp-supp:meander-eps0022}).\chgoff{}
\chgon This criterion detects the onset of splitting, whereas the definition of \cref{sec:assumptions} concerns the long-time evolution. In two dimensions the onset at one position need not lead to unbounded growth of $d$ there (\cref{sec:1d2d}), and the two-dimensional threshold fields reported below are thresholds of the onset.\chgoff{}
The threshold field is \chgfon searched for\chgoff{} by bisection in $|e|$, and each quoted threshold value is the midpoint of the final \chgfon bisection interval.\chgoff{}
}

\mk{mod}{2}{\Cref{tab:calc-params} lists the parameter values of the calculations.
For \GdFeO, first-principles exchange constants give $\varepsilon\approx0.02$, and $g\approx0.04$ is only an order-of-magnitude
estimate because the Fe anisotropy is borrowed from YFeO$_3$; the main calculations use $(\varepsilon,g)=(0.022,0.030)$,
consistent with these estimates within their \chgfon accuracy.\chgoff{}
The friction ratios $\gamma_n$ and $\gamma_P$ \chgon have not been measured and are inputs of the model. The ratio $\tau_P/\tau_N$ is expected to be below 1, and we vary it from $0.04$ to $40$.\chgoff{}
\chgon Over this range, the threshold field on the aligned side stays above the flat-wall threshold field, and the flat-wall threshold field stays above the threshold field on the opposed side (Supp.~\cref{supp-fig:gammaP}).\chgoff{}
\chgon The main calculations and the dependence on $g$ reported below use $\gamma_P=8g$ ($\tau_P/\tau_N=4$). The two-dimensional calculations were repeated at $\gamma_P=0.8g$ ($\tau_P/\tau_N=0.4$).\chgoff{}}
\mk{num}{2}{\chgfon Some of the calculations were repeated at $(\varepsilon,g)=(0.1,0.0512)$ with $\gamma_P=8g$. The crest-first splitting reported in \cref{sec:2d} was also found there,
and the one-dimensional interpretation of the two-dimensional threshold field also applies at this parameter set.\chgoff}
\mk{sup}{1}{\chgfon The numerical details
are described in Supp.~\cref{supp-supp:numerics},
and the calculations at $\varepsilon=0.1$ are reported in Supp.~\cref{supp-supp:series-eps01}.\chgoff}

\begin{table}[t]
\caption{Parameter values of the calculations. The main \chgfon parameter set\chgoff{} uses the values marked ``main'' and $\gamma_n=1$
($\varepsilon$ and $g$: \cref{eq:eps-g-def}).}
\label{tab:calc-params}
\begin{ruledtabular}
\begin{tabular}{lll}
Quantity & Value & Role \\ \hline
$\varepsilon=\tilde\kappa_G/\tilde\kappa_F$ & $0.022$ & main \\
$g$ & $0.030$ & main \\
 & $0.011$, $0.070$ & dependence on $g$ \\
$\gamma_n=\Gamma_n/\Gamma_N$ & $1$ & fixed \\
$\gamma_P=\Gamma_P/\Gamma_N$ & $8g$ & main \\
 & $0.08g$, $0.8g$, $80g$ & dependence on $\gamma_P$ \\
\end{tabular}
\end{ruledtabular}
\end{table}

\section{\label{sec:phase}One-Dimensional Results}

\mk{num}{2}{\chgon This section examines, by one-dimensional simulations of the TDGL equations, how the threshold field depends on the curvature (\cref{sec:main-result}).
Weak curvature raises the threshold field on the aligned side and lowers it on the opposed side.
\Cref{sec:theory-check} interprets this dependence as a balance between the difference of the drives on the constituent domain walls, which depends on the curvature, and the binding force between them.\chgoff{}
The results are numerical and hold for this model in one dimension at fixed curvature.}

\subsection{\label{sec:main-result}One-dimensional simulations: phase diagram}

\begin{figure*}[t]
\centering
\includegraphics[width=\linewidth]{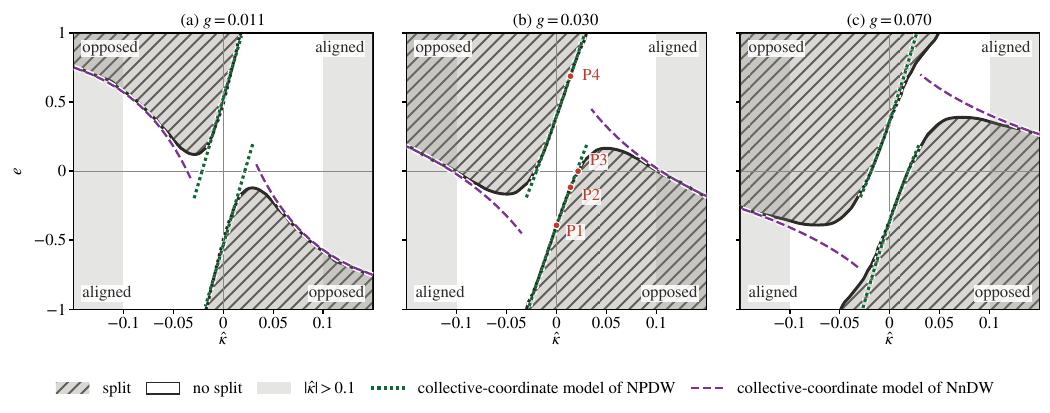}
\caption{Splitting region in the $(e,\hk)$ plane from one-dimensional calculations ($\varepsilon=0.022$, $\gamma_n=1$, $\gamma_P=8g$).
Solid line: threshold electric field $e_{\rm th}(\hk)$; \chgon dotted and dashed lines: collective-coordinate models of the \NPDW{} and of the \NnDW\chgoff{}
(\cref{sec:theory-check}), \chgfon with no fitted parameters.\chgoff{}
\chgon Each model line is drawn only over the range in which the model is compared with the numerical results.\chgoff{} Shaded: $|\hk|>0.1$, outside the \chgfon range of validity of the first-order approximation in curvature.\chgoff{}
\chgfon The points P1--P4 are defined in \cref{sec:main-result}.\chgoff}
\label{fig:phase}
\end{figure*}

\mk{num}{3}{For the \NPDW, we \chgfon start from a static wall ($\hk=0$, $e=0$) and, at $t=0$, switch on\chgoff{} a constant curvature $\hk$
and a \chgfon constant electric field $e$.\chgoff{}
We \chgfon scan\chgoff{} the $(e,\hk)$ plane in this way
for the three values of $g$ of \cref{tab:calc-params} at \chgfon$\gamma_P=8g$,
and determine whether the wall splits\chgoff{} (\cref{fig:phase}).}
\mk{num}{3}{By the symmetry of \cref{sec:model-props}, $(K,E)\to(-K,-E)$, the boundary in the $(e,\hk)$ plane is
point-symmetric about the origin: in each panel \chgfon of \cref{fig:phase}\chgoff, the two boundary lines separating the splitting from the non-splitting region
map onto each other.
We write $|e_{\rm th}(\hk)|$ for the \chgfon magnitude of the threshold field\chgoff{} at fixed curvature $\hk$,
the smallest $|e|$ that splits the wall on the side (aligned or opposed) considered.}

\mk{num}{3}{\chgfon In \cref{fig:phase}(b), the points P1 (flat wall), P2 (opposed side), and P4 (aligned side)
mark the three threshold fields compared in Result~1 below. The points P2 and P4 are at $|\hk|=0.014$.\chgoff}

\mk{num}{2}{For $g=0.030$ and $0.070$, the \chgfon threshold field on the opposed side\chgoff{} reaches zero at a finite curvature
[point P3 in \cref{fig:phase}(b)]\chgfon.
\chgoff{}\chgon{}Beyond this curvature, up to the limit $|\hk|=0.1$ of the weak-curvature approximation, the curvature drive alone splits the wall at $e=0$.\chgoff{}
\chgfon We call\chgoff{} this range of $|\hk|$ the curvature-only splitting range and write $|\hk_{\rm th}|$ for its lower end.
For $g=0.011$, the \chgfon threshold field on the opposed side\chgoff{} \chgon decreases as $|\hk|$ increases from zero and then\chgoff{} passes through a nonzero minimum\chgon. At this $g$ the wall therefore does not split by curvature alone at the curvatures examined.\chgoff}

\mk{num}{4}{Result~1:
\chgon For the parameter values examined here, weak curvature shifts the threshold field $|e_{\rm th}(\hk)|$ from its flat-wall value $|e_{\rm th}(0)|$,
upward on the aligned side
and downward on the opposed side.
Both shifts, each measured from $|e_{\rm th}(0)|$, grow as $|\hk|$ increases from zero.
For $g=0.030$ and $0.070$, these statements hold for $0<|\hk|<|\hk_{\rm th}|$, where curvature alone does not split the wall, and the threshold field on the opposed side vanishes at $|\hk_{\rm th}|$. For $g=0.011$, the statement for the opposed side holds for $|\hk|$ below the curvature of the minimum.\chgoff}

\subsection{\label{sec:theory-check}Interpretation in one dimension}

\mk{num}{2}{We interpret the one-dimensional threshold structure in \chgon three\chgoff{} steps.
\chgon In the plane of the field $e$ and the curvature $\hk$, a composite domain wall has a bound translating solution (\cref{sec:assumptions}) only inside a limited region.
We call the boundary of this region its existence limit.
First, we construct a collective-coordinate model of the \NPDW, which approximates the threshold curve of the \NPDW at small $|\hk|$.
The threshold is set by a balance between the difference of the drives on different constituent domain walls and the maximum restoring force of their binding.
Second, we ask whether the two composite domain walls produced by the splitting, the \NnDW and the \nPDW, have existence limits of their own, which bears on the shape of the threshold curve at larger $|\hk|$.
Third, we solve a boundary-value problem for the bound translating solution, which serves as a check on these limits and on the threshold curve at $|\hk|\lesssim0.1$.\chgoff}

\mk{num}{2}{\chgfon First, in the \NPDW on the\chgoff{}
narrow-$\Delta_G$ side, the relative angle $\psi$ turns from $0$ to $\pi$ in a region of width $\Delta_G$ near the Fe center,
where $p$ also
reverses. We therefore regard the \NPDW as a bound pair \chgfon of two composite domain walls,\chgoff{}\chgon{} an \NnDW-type wall at position $X_1$ and an \nPDW-type wall at position $X_2$.
Here $X_1$ and $X_2$ are the centers of the Fe and $P$ constituent domain walls, respectively.
Their distance $\Delta_r\equiv X_1-X_2$ is the counterpart, in the model, of the distance $d$ of \cref{sec:calc}.\chgoff{}
The binding energy $U_b(\Delta_r)$ is the energy of the wall with \chgon the two centers\chgoff{} held \chgfon a\chgoff{} distance
$\Delta_r$ apart, measured from its value at $\Delta_r\to\infty$\chgfon. It\chgoff{} is computed from the free energy of \cref{eq:F-dimless} at $e=0$ and $\hk=0$.
\chgfon The curvature and the field are treated, to first order, as forces acting on this static shape.
The thresholds therefore follow without solving the time evolution\chgoff{} (\cref{app:reduced-model}).
\chgon The resulting equation of motion for
$\Delta_r$ is\chgoff}
\begin{equation}
D_{\rm r}\,\frac{d\Delta_r}{dt}=F_b-U_b'(\Delta_r),\qquad
F_b=-\mathcal{A}\hk+\mathcal{B}e.
\label{eq:boundary-main}
\end{equation}
\mk{num}{2}{\chgfon Here\chgoff{} $F_b$ is the force conjugate to $\Delta_r$\chgon, $U_b'\equiv dU_b/d\Delta_r$ is the restoring force of the binding, and $D_{\rm r}$ is the dissipation of the relative motion (\cref{app:reduced-model}).\chgoff{}
A positive $F_b$ acts to increase $\Delta_r$.} \mk{num}{3}{\chgfon The term\chgoff{} $\mathcal{A}\hk$ is the relative curvature drive of
\cref{sec:curvature-drive}, which
remains because $\kappa_X/\Gamma_X$ differs among Fe, Gd, and $P$.
\chgfon The field acts only on $P$, and the term $\mathcal{B}e$ is the part of its force
that remains\chgoff{} after the common translation
of the pair is removed.} \mk{num}{2}{In the range of friction ratios and distances examined, $\mathcal{A},\mathcal{B}>0$.
\chgfon The two terms add on the opposed side
[$\hk e<0$, e.g., P2 in \cref{fig:phase}(b)] and partly cancel on the aligned side ($\hk e>0$, e.g., P4).\chgoff{}
\chgfon The maximum of the restoring force is\chgoff}
\begin{equation}
F_c\equiv\max_{\Delta_r}U_b'(\Delta_r).
\label{eq:Fc-def}
\end{equation}
\mk{num}{2}{The pair is stationary \chgon when the right-hand side of \cref{eq:boundary-main} equals zero, that is, $F_b=U_b'(\Delta_r)$. This equation has roots only for $|F_b|\le F_c$.
\chgoff{}\chgon{}For $0<|F_b|<F_c$ it has two roots.
The root at the smaller distance, where the restoring force increases with the distance, is the stable bound state, and the other root is unstable.\chgoff{}
\chgfon The wall splits when the two roots merge and disappear (a saddle-node bifurcation), at\chgoff}
\begin{equation}
|F_b|=F_c.
\label{eq:threshold-condition}
\end{equation}
\mk{num}{3}{\chgfon\Cref{eq:boundary-main,eq:threshold-condition} lead to the threshold field\chgoff}
\begin{equation}
|e_{\rm th}|=\ab(F_c\pm\mathcal{A}|\hk|)/\mathcal{B},
\label{eq:eth-linear}
\end{equation}
\mk{num}{3}{with $+$ on the aligned side and $-$ on the opposed \chgfon side,
provided that $\mathcal{A}$ and $\mathcal{B}$ are approximately independent of $\Delta_r$ at weak drive.
On both sides, $|e_{\rm th}|$ therefore varies linearly with $|\hk|$, with slopes of the same magnitude $\mathcal{A}/\mathcal{B}$ (the increase and the decrease of Result~1).
The threshold field on the opposed side\chgoff{} vanishes at $|\hk|=F_c/\mathcal{A}$,
the lower end of the curvature-only splitting \chgfon range.\chgoff{}
\chgon In the narrow-$\Delta_G$ limit this lower end has the leading term $\pi\varepsilon/4$ [\cref{eq:kappa-c-strong} in \cref{app:reduced-model}].
This value, in units of $1/\Delta_F$, serves as a guide to the curvature beyond which a cross section of a curved wall splits by curvature alone.\chgoff{}
\chgon The collective-coordinate model of the \NPDW, drawn as the dotted lines in \cref{fig:phase}, reproduces the threshold curve of the \NPDW at small $|\hk|$.\chgoff{}
\chgfon Because the model keeps the static shape computed at $e=0$ and $\hk=0$,
it applies only at weak drive
on the narrow-$\Delta_G$ side [$|\hk|\lesssim0.03$, $\Delta_G<\Delta_F$].\chgoff{}
\chgon For $g=0.030$, the intercepts of the model line with the two axes and its slope on the aligned side agree with those of the time-evolution threshold curve to within several percent of the latter (Supp.~\cref{supp-supp:intervals,supp-supp:steady-bvp}).\chgoff{}
\chgfon The derivation, including the $\Delta_r$ dependence of $\mathcal{A}$ and $\mathcal{B}$, is given in the \cref{app:reduced-model}.\chgoff}

\mk{num}{2}{\chgon Second, knowing that the splitting of the \NPDW produces an \NnDW and an \nPDW (\cref{sec:model-props}),
we expect that the \NPDW can split in this way only if both products exist as bound walls.
A collective-coordinate model of the \NnDW gives, for each field $e$, the largest curvature at which the \NnDW remains bound (Supp.~\cref{supp-supp:nndw-limit}). This curvature is drawn as the dashed lines in \cref{fig:phase}.
Beyond this curvature the model leaves the reversal of Gd behind the Fe constituent domain wall, so the splitting of the \NPDW cannot produce a bound \NnDW.
This limit is therefore a candidate for the phase boundary at larger curvature.
The last type of composite domain wall, the \nPDW, has no such limit in the range of \cref{fig:phase}.
For an isolated \nPDW to split, Fe would have to reverse, and the coupling torque on Fe in \cref{eq:tdgl-dimless}, at most $2g(1+|e|)$, stays below the largest anisotropy torque $1/2$ throughout this range.\chgoff}

\mk{num}{2}{\chgfon For $g=0.011$ and $0.030$, as $|\hk|$ increases beyond $|\hk|\approx0.03$ the boundary approaches the dashed lines.
We conjecture that the boundary crosses over from the dotted line to the dashed line.\chgoff{}
\chgon The two lines together would then roughly account for the curvature at which the threshold field turns, which gives the threshold curve its peaked shape around $|\hk|\approx0.03$ to $0.05$.\chgoff}

\mk{num}{2}{\chgon Third, as a check, we\chgoff{} solve a boundary-value problem for a bound translating solution at constant \chgfon curvature.\chgoff{}
\chgon The problem is the steady form of \cref{eq:tdgl-dimless} in the frame that moves with the wall at an unknown common velocity, and we follow its solution branch numerically by Newton's method with continuation in the drive.
For the \NnDW, the existence limit obtained in this way agrees with the dashed lines of \cref{fig:phase}, over the range drawn, to within $5\%$ of the former (Supp.~\cref{supp-supp:steady-bvp}).\chgoff{}
\chgfon For the \NPDW, as $|e|$ is increased, the bound translating solution disappears at a turning point (fold) of the solution branch\chgoff{}\chgon,
as the steady front does in models with two nonconserved order parameters~\cite{TuckermanBechhoefer1992}.
\chgfon For all $g$ and $\varepsilon$ examined\chgoff{}\chgon{} at $|\hk|\lesssim0.1$\chgoff{}\chgfon{}, the position of this fold\chgoff{}\chgon{} coincides with the threshold $|e_{\rm th}(\hk)|$ shown in \cref{fig:phase}.\chgoff{} \chgfon The threshold thus marks\chgoff{} the loss of the bound translating solution.}

\mk{sup}{1}{\chgfon The threshold intervals of the time evolution are given in Supp.~\cref{supp-supp:intervals},
the formulation of the boundary-value problem in Supp.~\cref{supp-supp:bvp-method},
and the comparisons at larger curvature in Supp.~\cref{supp-supp:steady-bvp}.\chgoff}

\section{\label{sec:2d}Two-Dimensional Results and Their Interpretation \texorpdfstring{\chgfon Based on the\chgoff}{Based on the} One-Dimensional Results}

\mk{gen}{2}{The results so far concern a one-dimensional wall at a given curvature.}
\mk{num}{3}{For a meandering wall \chgfon under\chgoff{} a uniform field, the phase diagram of \cref{sec:main-result},
\chgfon evaluated at the curvature of each apex, suggests where the wall splits.
Suppose that $|e|$ lies between $|e_{\rm th}|$ on the opposed side and $|e_{\rm th}|$ on the aligned side at the apex curvature.
Then $|e|$ is above the threshold field at the crest (opposed) and below it at the trough (aligned),
so\chgoff{} the splitting should start at the crest.}
\mk{num}{2}{\Cref{sec:1d2d} tests this expectation with
two-dimensional calculations of the same model, in which the curvature arises from the wall shape and changes in \chgon time.
In these calculations a meandering wall splits first at the crest, as expected.
\Cref{sec:2d-interp}
interprets the two-dimensional threshold field with one-dimensional calculations.
Below the lower end of the curvature-only splitting range, the decay of the apex curvature accounts for most of the difference between the threshold field found in the two-dimensional simulations and the one-dimensional one at fixed curvature.\chgoff}

\subsection{\label{sec:1d2d}Two-dimensional simulations: meandering wall}

\mk{mod}{2}{The initial \chgfon \NPDW\chgoff{} is \chgfon set as\chgoff{} a sinusoid $x=w\sin(2\pi y/L_y)$,
where $x$ is normal to the wall and $y$ is along the \chgfon wall.\chgoff{}
\chgfon Since the field drives the wall in the $+x$ direction,
the crest is at $y=L_y/4$ and the trough at $y=3L_y/4$.
The curvature $\hk$ of the wall depends on time and on $y$.
We write $\hk_{\rm v}(t)$ for its value at the crest (the crest apex curvature),
with the sign chosen so that $\hk_{\rm v}$ is positive for the initial bending of the crest (opposed, $\hk e<0$).
The initial value is $\hk_{\rm v}(0)=wk^2\Delta_F$ with $k=2\pi/L_y$.\chgoff}

\begin{figure*}[t]
\centering
\includegraphics{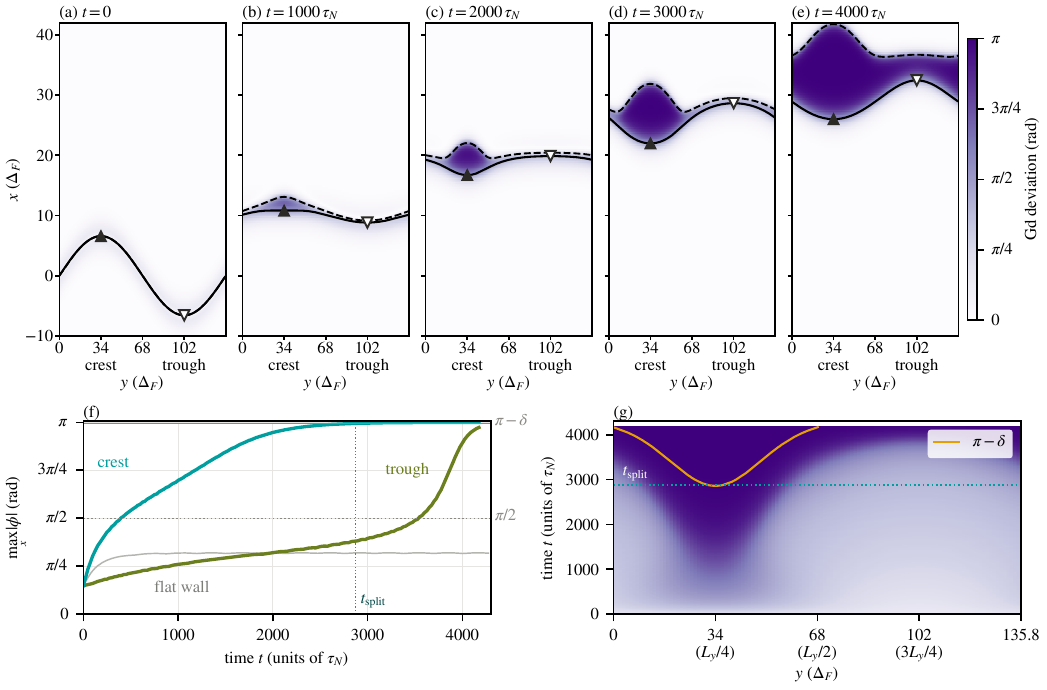}
\caption{Splitting of a two-dimensional meandering wall at the crest [$(\varepsilon,g)=(0.022,0.030)$, $\gamma_P=8g$,
$L_y=135.8\,\Delta_F$, $|\hk_{\rm v}(0)|=0.0141$, $e=-0.325$].
(a)--(e) Gd deviation \chgfon in the plane of $y$ and $x$ at five times\chgoff{} ($x=0$: mean Fe wall position at $t=0$); solid and dashed lines: Fe and
$P$ centers.
(f) \chgfon Time dependence of $\max_x|\phi|$ on the line cuts along $x$ through the crest and through the trough.\chgoff{} (g) $\max_x|\phi|$ \chgfon as a function of $y$ and $t$;\chgoff{} contour: $\pi-\delta$.}
\label{fig:2d}
\end{figure*}

\mk{num}{3}{\Cref{fig:2d} shows a meandering wall driven at $e=-0.325$ (parameters in the caption),
a field at which a flat wall does not split [\cref{fig:2d}(f)].
A region of reversed Gd, i.e., the third bulk state, appears at the crest
and spreads along the wall into the trough half \chgfon of the computational domain ($L_y/2\le y<L_y$)\chgoff{} [\cref{fig:2d}(a)--(e)].
\chgfon The splitting criterion of \cref{sec:calc} shows the same process.
On the line cut along $x$ through the crest ($y=L_y/4$),\chgoff{}
the Gd deviation reaches $\pi-\delta$, whereas \chgfon on the line cut through the trough\chgoff{} ($y=3L_y/4$) it stays below
$\pi-\delta$ until the end of the run [\cref{fig:2d}(f)].
\chgfon The\chgoff{} region where the Gd deviation reaches $\pi-\delta$ starts at the crest
and extends along the \chgfon wall, and every part of the trough half that reaches $\pi-\delta$ is connected to this region [\cref{fig:2d}(g)].\chgoff{}
\chgon \chgfon We examined eight initial apex curvatures. In every two-dimensional run in which the wall split, with one exception described below, the splitting started at the crest.\chgoff{}
\chgon The crest also splits first at $\gamma_P=0.8g$, at fields at which a flat wall does not split at this $\gamma_P$ (Supp.~\cref{supp-supp:meander-eps0022}).\chgoff}

\mk{num}{3}{\chgfon The crest splitting of \cref{fig:2d} is caused by curvature and field together.
The initial apex curvature of this wall\chgoff{} is smaller than the lower end of the curvature-only splitting range
obtained in one dimension at constant curvature, and \chgon after the initial relaxation\chgoff{} the apex curvature \chgfon never increased.\chgoff{}
The crest nevertheless split, \chgfon at a threshold field $|e_{\rm th}|$ below the flat-wall $|e_{\rm th}|$.\chgoff}

\mk{num}{3}{\chgon Near the threshold fields, the two-dimensional runs showed two exceptional behaviors, presumably caused by effects specific to two dimensions.
First, a region of the third bulk state that has formed at the crest can start to close again.
This occurred in some of the runs at the smallest fields at which the crest met the $\pi-\delta$ criterion.
The tension from the adjoining unsplit part of the wall is the likely cause (Supp.~\cref{supp-supp:split-close}).
The Gd deviation at the crest once exceeded $\pi-\delta$ but later fell below it again, and the distance $d$ at the crest was decreasing when these runs ended, so the onset is most likely not followed by splitting in the sense of \cref{sec:assumptions}.
The closing starts after the onset and leaves the position of the onset unchanged.\chgoff}

\mk{num}{3}{\chgon Second, one run is the exception to the crest-first order noted above.
In this run, at a field slightly above the flat-wall threshold field, the meander relaxed by a damped oscillation.
During this oscillation, the bending of the \NPDW reversed and returned before the wall split, and $\pi-\delta$ was first reached at the initial trough.
The calculations of Supp.~\cref{supp-supp:reversal} indicate that this position was selected while the bending was reversed, when it was the apex protruding in the direction of motion.
This exception could therefore result from the timing of the splitting inside the wall relative to the damped oscillation of the whole wall.\chgoff}

\mk{num}{4}{We summarize \chgfon the results of this subsection.\chgoff{}\chgon{} They were obtained at the main parameter set for $\gamma_P=8g$ and $0.8g$,
and they apply to walls whose bending keeps its initial orientation
until the wall splits.\chgoff{}
Result~2: A meandering wall driven by a uniform field splits first at the crest, and the split region
then extends along the wall toward the trough\chgfon. The\chgoff{} splitting does not start at the trough
\chgfon(interpreted in \cref{sec:2d-interp})\chgoff.
\chgon Immediately above the two-dimensional threshold field, however, the split region can start to close again, presumably through the tension from the adjoining unsplit part of the wall.\chgoff}
\mk{sup}{1}{\chgfon Details of the runs, including those at $\gamma_P=0.8g$, are given in Supp.~\cref{supp-supp:meander-eps0022}.\chgoff}

\subsection{\label{sec:2d-interp}Interpretation in two dimensions}

\begin{figure}[t]
\centering
\includegraphics{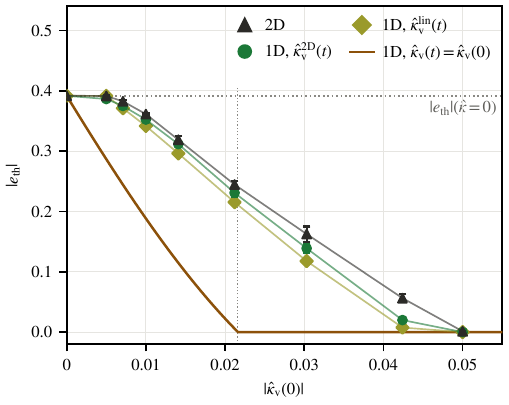}
\caption{\chgfon Threshold electric field for splitting at the crest (opposed) versus the initial apex curvature\chgoff{}
[$(\varepsilon,g)=(0.022,0.030)$, $\gamma_P=8g$, $L_y=135.8\,\Delta_F$; series: \cref{sec:2d-interp}].
Line $\hk_{\rm v}(t)=\hk_{\rm v}(0)$: boundary-value problem (Supp.~\cref{supp-supp:bvp-method}), on which the time-evolution thresholds of
\cref{fig:phase}(b) lie.}
\label{fig:meander-average}
\end{figure}

\mk{num}{2}{We interpret \chgfon the threshold field for the onset of splitting at the crest (triangles in \cref{fig:meander-average})\chgoff{} with one-dimensional calculations.
At the same initial apex curvature, \chgfon the threshold field $|e_{\rm th}|$ at the crest of the two-dimensional wall differs from
$|e_{\rm th}|$\chgoff{} of a one-dimensional wall at fixed curvature.
\chgon Two factors are expected to contribute to this difference. One is the time dependence of the curvature (temporal effect). The meandering wall flattens during the run, so its apex curvature decays (\cref{sec:1d2d}). The other is the coupling between neighboring parts of the wall through the stiffness terms along the wall (spatial effect), which the one-dimensional calculations omit. We examine how much of the difference the temporal effect explains.\chgoff{}
\Cref{fig:meander-average}
compares the crest $|e_{\rm th}|$ of the two-dimensional meandering wall (triangles) with three
\chgfon series of one-dimensional calculations. The series differ only in the time dependence of the signed apex curvature $\hk_{\rm v}(t)$ used as input.
The first uses $\hk_{\rm v}^{\rm 2D}(t)$, the curvature measured in the two-dimensional run (circles).\chgoff{}
\chgfon The second uses $\hk_{\rm v}^{\rm lin}(t)\equiv k^2x_{\rm Fe}(t)\,\Delta_F$ (diamonds), where\chgoff{}
$x_{\rm Fe}(t)$ is the crest displacement of the Fe constituent domain wall center, obtained \chgfon theoretically\chgoff{} from the linear modes of the
flat wall translating at the same \chgfon field. The third keeps the curvature fixed at $\hk_{\rm v}(0)$ (line), which corresponds to the threshold curve on the opposed side of \cref{fig:phase}(b).\chgoff{}
\chgfon Each series applies the same curvature to all order parameters.\chgoff{}
\chgon This is an approximation, because in the two-dimensional wall the curvature differs between the constituent domain walls.
The center line of Fe, the stiffest order\chgfon{} parameter\chgoff{}\chgon, flattens earlier than that of $P$ [\cref{fig:2d}(b)].\chgoff{}
\chgfon The one-dimensional series use the apex curvature of the Fe center line.\chgoff}

\mk{num}{2}{Below the lower end of the curvature-only splitting range
[vertical dotted line; the range that starts at point P3 in \cref{fig:phase}(b)],
the fixed-curvature threshold field lies below the two-dimensional one at every point examined (\cref{fig:meander-average})\chgfon.\chgoff{}
\chgon After the initial relaxation, the apex curvature in the two-dimensional runs of the figure decays, and the curvature drive weakens with it.
The fixed-curvature series, which keeps the initial strength of the curvature drive, therefore gives the lowest threshold field $|e_{\rm th}|$ of the three one-dimensional series.\chgoff}
\mk{num}{3}{In the same range, \chgfon the difference in $|e_{\rm th}|$ between each series with a time-dependent curvature and the two-dimensional simulation result (triangles)
is an order of magnitude smaller than the difference between the fixed-curvature series and the triangles.\chgoff{}
\chgon The temporal effect therefore accounts for most of the difference between the two-dimensional and the fixed-curvature threshold fields at $|\hk_{\rm v}(0)|<|\hk_{\rm th}|$.\chgoff{}
Since $\hk_{\rm v}^{\rm lin}(t)$ \chgfon is obtained without any two-dimensional simulation, this interpretation\chgoff{} of the crest threshold field rests on two
one-dimensional ingredients: the decay of the meander given by the linear modes of the flat wall, and the response of the
one-dimensional wall to the resulting \chgfon time dependence of the curvature\chgoff.}
\mk{num}{2}{Inside the curvature-only splitting range, the fixed-curvature threshold field is $0$, whereas the
two-dimensional wall does not split by curvature alone at the points \chgfon examined.
At $|\hk_{\rm v}(0)|=0.0303$, the two series with a time-dependent curvature lie further below the triangles in $|e_{\rm th}|$\chgoff{}
than at smaller \chgfon curvatures\chgoff{} (\cref{fig:meander-average}).}

\mk{num}{2}{\chgon Below the lower end of the curvature-only splitting range, the one-dimensional $|e_{\rm th}|$ of the trough (aligned), taken at the initial apex curvature, exceeds
its flat-wall value (Result~1).
A field below the flat-wall $|e_{\rm th}|$, as in the run of \cref{fig:2d}, is therefore also below this trough $|e_{\rm th}|$,
and the one-dimensional interpretation\chgoff{}
predicts no trough-first splitting.
\chgon This prediction uses the initial apex curvature, whereas the curvature of the two-dimensional wall decays. We therefore also drove the one-dimensional wall with a trough apex curvature that decays in time, \chgfon taken either from the two-dimensional run or as the exponential decay of the curvature given by the relaxation of the wall shape.
At\chgoff{}\chgon{} the smallest $|e|$ at which the two-dimensional wall split, \chgfon this wall stayed bound\chgoff{}\chgon{}
for $|\hk_{\rm v}(0)|$ from $0.0071$ to $0.0424$, a range that extends above the lower end of the curvature-only splitting range.\chgoff}

\mk{sup}{1}{The remaining \chgfon difference in $|e_{\rm th}|$ between the triangles and the circles\chgoff{}
may include the spatial \chgfon effect.
The construction of the curvature time series and the trough-side calculations are given in Supp.~\cref{supp-supp:curvature-series},
and the comparison with circular domains in Supp.~\cref{supp-supp:series-eps01}.\chgoff}

\section{\label{sec:discussion}Discussion}

\mk{con}{3}{The crest--trough selection \chgfon of Result~2 can be tested in GdFeO$_3$\chgoff{} by driving a meandering wall with a uniform field and observing
whether splitting
starts at the crest or at the trough\chgfon. The\chgoff{} start is marked by a band of the third bulk state between the Fe (magnetization)
constituent domain wall and the $P$ (polarization) constituent domain wall.}
\mk{con}{1}{For \chgfon this material\chgoff{} (Supp.~\cref{supp-supp:params}),
the curvatures of Result~1 correspond to radii of curvature of order $1\,\muup$m, and the
flat-wall threshold \chgfon electric\chgoff{} field is of order $10$\,kV/cm.}
\mk{con}{2}{Possible probes are magneto-optical Kerr effect (MOKE) microscopy via the weak ferromagnetism of Fe, and
second-harmonic generation and piezoresponse force microscopy for the polarization\chgfon.
The measurements must be made at\chgoff{} temperatures below the Néel temperature of \chgfon Gd, $2.5$\,K~\cite{Tokunaga2009}.\chgoff{}
\chgon Imaging at such low temperatures may make the experiment difficult in practice.\chgoff}

\mk{con}{2}{\chgon The curvature dependence of splitting, the central result of this paper, may also be tested in other multi-order systems
that meet two necessary conditions.
\chgoff{}\chgon One is that the stiffness-to-friction ratio $\kappa_X/\Gamma_X$ differs among the order parameters, so that \cref{eq:prop1} is violated.
\chgfon The other is the existence of a locally stable third bulk state, which fills the region between the two separated composite domain walls (\cref{sec:assumptions}).\chgoff{}
Splitting itself requires this state, by the free-energy argument of that subsection.
Note that meeting these two conditions does not by itself guarantee that the wall splits, or that crest and trough split at different threshold fields.\chgoff{}\chgfon{}
\chgoff{}\chgon The binding of the constituent domain walls can act as an energy barrier that still prevents the splitting.
A field that drives all the constituent domain walls together exerts no relative drive, so that curvature and field need not compete.\chgoff}

\mk{con}{2}{\chgon Outside bulk multiferroics, three classes of systems are candidates for testing the curvature dependence of splitting.
The first is a reaction--diffusion system
$\partial_t\phi_X=(\kappa_X/\Gamma_X)\nabla^2\phi_X+R_X(\phi_1,\dots,\phi_n)$
with unequal diffusion coefficients $\kappa_X/\Gamma_X$, which meets the condition by construction.
The reaction terms of a general reaction--diffusion system need not derive from a free energy, and Proposition~1 and Corollary~1 do not require them to.
Both therefore also hold for such a system with constant diffusion coefficients, to first order in weak curvature.\chgoff{}
\chgfon The second is a magnetic multilayer\chgoff{} whose ferromagnetic layers are coupled across a nonmagnetic spacer,
in which bound and separate motion of the walls of two layers \chgon of different thickness\chgoff{} has been observed~\cite{Metaxas2010}.
\chgon Layers that differ in thickness or material are known to differ in domain-wall mobility.
A test requires that each wall move steadily, below its Walker field and above its depinning field,
and that stray fields be weak compared with the interlayer coupling.
The third is a liquid crystal whose orientational order is coupled to a second nonconserved order of a different physical origin,
so that the ratios $\kappa_X/\Gamma_X$ of the two orders can be expected to differ.
A possible example is a ferronematic, a suspension of magnetic nanoparticles in a nematic liquid crystal,
in which domains of opposite magnetization form on cooling without a magnetic field~\cite{Mertelj2013}.
Its nematic order and magnetization have been modeled by coupled relaxational equations with separate elastic and damping constants~\cite{Vats2021},
and liquid-crystal patterns are generally large enough to be observed optically.\chgoff}

\mk{con}{2}{\chgon This study has the following limitations.
All results assume noise-free relaxational dynamics of nonconserved order parameters without disorder,
and the numerical results are only for the \GdFeO-type model, in the parameter ranges of \cref{sec:calc}.
The analytical results hold to first order in weak curvature, and Corollary~1 concerns one-dimensional walls.
As a description of a magnet, the model neglects stray fields and the precessional dynamics of the spins.
Experimental thresholds also depend on coercivity and pinning.
The site at which the wall first splits (crest or trough) is therefore better suited to a comparison with experiment than the value of the threshold electric field.\chgoff}

\section{\label{sec:conclusion}Conclusion}

\mk{con}{3}{\chgon We investigated how wall curvature changes the threshold field for the splitting of a composite domain wall
and how it selects where the wall splits.
We took a Ginzburg--Landau model of \GdFeO{} type as an example of a system with several coupled order parameters.\chgoff{}}

\mk{num}{3}{\chgon We interpret both effects \chgfon of curvature\chgoff{}\chgon{} by one mechanism.
In a curved wall, once the common motion of the whole wall is removed, curvature leaves a relative drive between its constituent domain walls, and the field exerts \chgfon a relative drive\chgoff{}\chgon{} of its own.
The two relative drives add when curvature pushes the wall against the field,
which lowers the threshold field magnitude below its flat-wall value.
They partly cancel when curvature pushes the wall along the field, which raises \chgfon this magnitude above the flat-wall value.\chgoff{}\chgon{}
One-dimensional calculations with a weak curvature held fixed gave this lowering and raising.
In these calculations the threshold field coincides with the largest field at which the wall has a bound translating solution.
An analysis using the collective-coordinate model accounts for this mechanism at weak drive.
\chgoff{}}

\mk{num}{3}{\chgon The same mechanism suggests where a curved wall splits.
Under a uniform field, the two relative drives add where the wall bulges forward along the field-driven motion,
so this part should split first.
In two-dimensional calculations of a meandering wall in this model, the wall indeed split first at the forward apex (the crest).
Where the initial apex curvature is too weak to split the one-dimensional wall without a field, the threshold field at the crest differs from the one-dimensional value at fixed curvature mostly because the meander flattens before the wall splits.
Just above this threshold field, the split region can start to close again, presumably through the tension from the neighboring unsplit part of the wall.\chgoff{}}

\mk{gen}{4}{\chgon Our analysis shows what this mechanism requires.
It requires (i) a ratio of stiffness to friction that differs between order parameters, for curvature to change the threshold field,
and (ii) a locally stable third bulk state, for the wall to split at all.
Condition (i) is derived for noise-free relaxational dynamics of nonconserved order parameters at linear order in curvature.
There, curvature leaves a relative drive exactly when these ratios are unequal,
\chgfon and it can change the threshold field for splitting of a one-dimensional composite domain wall only then.\chgoff{}
Both conditions are necessary ones and involve only the local free energy, the stiffnesses, and the frictions.
\chgoff{}}

\begin{acknowledgments}
\chgon The authors thank Hirokazu Tsunetsugu for \chgfon suggesting the study of\chgoff{}\chgon{} systems with multiple order parameters and for discussions on driving by an electric field.\chgoff{}
\end{acknowledgments}

\appendix
\crefalias{section}{appendix}

\section{\label{app:reduced-model}\texorpdfstring{Coefficients and Thresholds of the Collective-Coordinate Model \chgfon of the \NPDW\chgoff}{Coefficients and Thresholds of the Collective-Coordinate Model of the NPDW}}

\mk{app}{2}{The coefficients and thresholds of the collective-coordinate model \chgon of the \NPDW (\cref{sec:theory-check})\chgoff{} follow from
\chgfon the easy-plane equations of motion, \cref{eq:tdgl-dimless}, alone. Their free-energy density is\chgoff}
\begin{equation}
\begin{aligned}
\frac{f_{\rm tot}}{\tilde\alpha_x}&=\frac12(\partial_z\Phi)^2+\frac{\varepsilon}{2}(\partial_z\phi)^2-\frac12\cos^2\Phi\\
&\quad+g\,p^2-2g\,e\,p+2g\,p\cos\psi .
\end{aligned}
\label{eq:F-dimless}
\end{equation}

\mk{app}{2}{
\chgfon The model rests on four approximations. (i) The Fe profile is fixed to that of the
uncoupled static Fe wall, $\Phi_0(z-X_1)$ ($\hk=0$, $e=0$).
(ii) The polarization follows the relative angle instantaneously, $p=e-\cos\psi$, and
$\phi=\Phi-\psi$. (iii) For each $\Delta_r=X_1-X_2$,
the shape $\psi=\Psi(z-X_2;\Delta_r)$ is obtained numerically from a static boundary-value
problem at $e=0$ and $\hk=0$, with the center of $\psi$ held at $X_2$.
(iv) The field and the curvature enter only as forces projected onto that
shape, to first order. The binding energy\chgoff{} $U_b(\Delta_r)$ is the minimum, at fixed $\Delta_r$, of the $\psi$-dependent part of \cref{eq:F-dimless},
measured from its value at
$\Delta_r\to\infty$.}

\mk{app}{2}{The friction weights $(1,\gamma_n,\gamma_P)$ and the curvature weights
$(1,\varepsilon,0)$ of the three equations define the dissipation matrix and the curvature-weight matrix,}
\begin{equation}
\begin{aligned}
D_{ij}&=\int dz\,\ab(\partial_{X_i}\Phi\,\partial_{X_j}\Phi+\gamma_n\,\partial_{X_i}\phi\,\partial_{X_j}\phi+\gamma_P\,\partial_{X_i} p\,\partial_{X_j} p),\\
\Lambda_{ij}&=\int dz\,\ab(\partial_{X_i}\Phi\,\partial_{X_j}\Phi+\varepsilon\,\partial_{X_i}\phi\,\partial_{X_j}\phi).
\end{aligned}
\label{eq:rm-DG}
\end{equation}
\mk{app}{2}{The forces on $X_1$ and $X_2$, positive in the $-\xi$ direction, are}
\begin{equation}
F_i=\hk\sum_j\Lambda_{ij}-2ge\int dz\,\sin\Psi\,\partial_{X_i}\psi,
\label{eq:rm-forces}
\end{equation}
\mk{app}{2}{\chgfon whose\chgoff{} field term sums to $4ge$ over $i=1,2$.
Eliminating the center-of-mass
motion from the overdamped equations \chgon$\sum_jD_{ij}\dot X_j=-F_i-\partial_{X_i}U_b$\chgoff{} \chgfon gives the equation for the distance $\Delta_r$,\chgoff}
\begin{equation}
\begin{gathered}
D_{\rm r}\,\dot\Delta_r=F_b-U_b'(\Delta_r),\\
F_b=\frac{F_2(D_{11}+D_{12})-F_1(D_{22}+D_{12})}{D_{11}+D_{22}+2D_{12}},\\
\chgfon D_{\rm r}=\frac{D_{11}D_{22}-D_{12}^2}{D_{11}+D_{22}+2D_{12}},\chgoff{}
\end{gathered}
\label{eq:collective-coordinate}
\end{equation}
\mk{app}{2}{where \chgfon$D_{\rm r}$\chgoff{} is the dissipation of the relative motion.
\chgfon The force\chgoff{} $F_b$ has the form of
\cref{eq:boundary-main}, with $\Delta_r$-dependent $\mathcal{A}$ (from $\Lambda_{ij}$ and $D_{ij}$) and $\mathcal{B}$ (from
$D_{ij}$ and the
field term). \chgfon It is unchanged by a common translation of the pair, i.e., by a force proportional to $\sum_jD_{ij}$ added to every $F_i$,
and is therefore the relative drive.\chgoff{}
For $\varepsilon=\gamma_n$ and $\gamma_P=0$, $\Lambda_{ij}=D_{ij}$, so the curvature force \chgfon produces only\chgoff{} such
a common translation and $\mathcal{A}=0$, as Proposition 1 [\cref{eq:prop1}] requires.
For \chgfon$g=0.030$,\chgoff{} $\mathcal{A},\mathcal{B}>0$ over the ranges of $\Delta_r$ and friction ratios \chgfon examined. Outside these ranges, $\mathcal{B}$ becomes negative as $\Delta_r\to0$ for
$\gamma_n\gtrsim3.2$.\chgoff}

\mk{app}{2}{The thresholds are static quantities. The wall splits when \cref{eq:collective-coordinate} loses its stationary root
$F_b=U_b'(\Delta_r)$; for $\Delta_r$-independent $\mathcal{A}$ and $\mathcal{B}$ this is \cref{eq:threshold-condition}.
\chgfon When the $\Delta_r$ dependence of
$\mathcal{A}$ and $\mathcal{B}$ is included,\chgoff{} the lower end of the curvature-only splitting range is $\max_{\Delta_r}\ab(U_b'/\mathcal{A})$ at
$e=0$, the
flat-wall threshold field is $\max_{\Delta_r}\ab(U_b'/\mathcal{B})$ at $\hk=0$,
and the slope of the threshold curve on the aligned side is
$\mathcal{A}/\mathcal{B}$ at the $\Delta_r$ that maximizes $U_b'/\mathcal{B}$.
\chgfon The dissipation\chgoff{} $D_{\rm r}$ sets only the time scale of the motion of $\Delta_r$
and does not enter the \chgfon thresholds.\chgoff{}
In the narrow-$\Delta_G$ limit, $\Delta_G/\Delta_F\to0$ with $\varepsilon(1+\gamma_n)\ll\gamma_n+\frac23\gamma_P$,
\chgfon the lower end becomes\chgoff}
\begin{equation}
|\hk_{\rm th}^{\rm cc}|\approx\frac{\pi\varepsilon}{4}.
\label{eq:kappa-c-strong}
\end{equation}
\mk{app}{2}{Numerical evaluation for $\varepsilon=0.022$, $g=0.030$, $\gamma_n=1$,
$\gamma_P=8g$ \chgfon provides\chgoff{} the values compared in \cref{sec:theory-check}.
\chgfon The static-shape equation, the narrow-$\Delta_G$ limit, the numerical grids, and the examined ranges of $\Delta_r$ and friction ratios
are documented in Supp.~\cref{supp-supp:cc-range}.\chgoff}

\bibliographystyle{jpsj_rules2024}
\bibliography{GdFeO3_curvedDW_jpsj,main_jpsj_extra}%

\end{document}